\documentclass[prb,fleqn,twocolumn, prb, superscriptaddress]{revtex4}
\usepackage{epsfig}
\usepackage{graphicx}
\usepackage{color} 
\usepackage{amsfonts}

\newcommand{\ct}{\cite}
\newcommand{\bi}{\bibitem}
\newcommand{\be}{\begin{equation}}
\newcommand{\ee}{\end{equation}}
\newcommand{\ba}{\begin{eqnarray}}
\newcommand{\ea}{\end{eqnarray}}
\newcommand{\al}{\alpha}

\newcommand{\non}{\nonumber}
\newcommand{\bra}[1]{\langle #1|}
\newcommand{\ket}[1]{|#1\rangle}
\newcommand{\de}{\delta}

\begin{document}
\title{Study of Loschmidt Echo for two-dimensional Kitaev model}
\author{Shraddha Sharma}
\email{shrdha@iitk.ac.in}
\affiliation{Department of Physics, Indian Institute of Technology Kanpur, 
Kanpur 208016, India}
\author{Atanu Rajak}
\email{atanu.rajak@saha.ac.in}
\affiliation{TCMP Division, Saha Institute of Nuclear Physics, 1/AF Bidhannagar, Kolkata 700 064, India}

\begin{abstract}
In this paper, we study the Loschmidt Echo (LE) of  a two-dimensional Kitaev model residing on a honeycomb lattice which is chosen to be an environment that is 
coupled globally to a central spin. The 
decay of LE is highly influenced by the quantum criticality of the environmental spin model e.g., it shows a sharp dip close to the  
anisotropic quantum critical point (AQCP) of its phase diagram. The early time decay and the collapse and revival as a function of time at AQCP do also
 exhibit interesting
 scaling behavior with the system size which is verified numerically. It has also been observed that the LE stays 
vanishingly small throughout the gapless
 phase of the model. The above study has also been extended to the 1D Kitaev model i.e. when one of the interaction terms vanishes. 
\end{abstract}

\maketitle   
\section{Introduction}
\label{I}
A quantum phase transition is a zero temperature transition of a quantum many body system driven by a non-commuting term of the quantum Hamiltonian
which is associated with a diverging length as well as a diverging time scale \ct{sachdev99,chakrabarti96}. In recent years, a
 plethora of studies are being carried out which attempt to bridge a connection
between quantum phase transition and quantum information theory \ct{nielsen00,vedral07}. For example, information theoritic measures like  entanglement, quantum
 fidelity \ct{zanardi06,zhou08,gu10,gritsev10}, decoherence \ct{zurek91,haroche98,zurek03,joos03} 
and quantum discord \ct{sarandy09,dillenschneider08}, etc., 
are being studied close to the quantum critical point (QCP). These measures not only capture the singularities associated
with the QCP but also show distinct scaling relations which characterizes it. There have also been numerous studies on decoherence (or loss of phase information) in
 a quantum critical system which is closely connected to the LE to be discussed in this work; understanding decoherence is essential for successful achievement 
of the quantum computation.
 
To study the LE in a quantum critical environment, we make resort to the 
central spin model \ct{zanardi} in which a central spin $S$ is coupled globally to an environmental spin model $E$ (which in this case is the 
two dimensional Kitaev model).
 The LE (with the $E$ in some ground state $|\psi_0\rangle$) is given by
$$ L(t) = |\langle \psi_0| e^{iH_0t} e^{-i(H_0 + \de H_\de)t} |\psi_0\rangle|^2. $$
Here the $H_0$ and $H_0+\de H_\de$ are the two Hamiltonians with which the ground state $|\psi_0\rangle$ evolves, where the term $\de H_\de$
 arises due to the coupling of the $E$ with the $S$.
It has been established that the LE shows a decay near
 the critical point of the $E$ with a decay rate that marks the universality associated with the QCP of $E$ \ct{zanardi, zhang09, rossini07, sharma12}. Also 
the LE shows collapse
and revival as a function of time when the $E$ is at the QCP.

The proposed work is organized in the following way: Sec.\ref{I} presents the model Hamiltonian, the phase diagram and discussion about 
the AQCP. In sec.\ref{II}, we describe
the general calculation of the LE and in the subsequent subsections we study the scaling of the 
short time decay close to the AQCP and its collapse and revival with time.

\section{Model, Phase Diagram and Anisotropic Quantum Critical Point (AQCP)}
\label{II}
\begin{figure}[ht]
\begin{center}
\includegraphics[width=7.7cm]{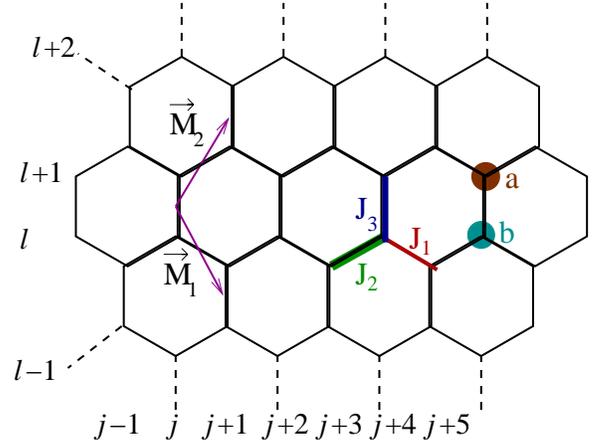}
\end{center}
\caption{(Color online) Kitaev model on a honeycomb lattice with $\vec{M_1}$ and $\vec{M_2}$ being spanning 
vectors of the lattice and $J_1$, $J_2$ and $J_3$, the coupling on the three bonds [\onlinecite{kitaev06}].}
\label{Fig1}
\end{figure}

The Hamiltonian of the Kitaev model on a honeycomb lattice is given by
\be
H=\sum_{j+l=even}\left(J_{1}\sigma^{x}_{j,l}\sigma^{x}_{j+1,l}+J_{2}\sigma^{y}_{j-1,l}\sigma^{y}_{j,l}+J_{3}\sigma^{z}_{j,l}\sigma^{z}_{j,l+1}\right)
\label{ham1}
\ee
where $j$ and $l$ signify the column and row indices respectively of the honeycomb lattice  while  $J_1$, $J_2$ and $J_3$ are 
coupling parameters for the three bonds (see Fig.~(\ref{Fig1})) \ct{kitaev06, sengupta08}; and $\sigma^{\al}_{j,l}$,  are the Pauli spin matrices 
with $\al$($=x$, $y$ and $z$), denoting the spin component.

We will assume the parameters $J_1$, $J_2$ and $J_3$ are all positive and 
confine our analysis on the plane $J_1+J_2+J_3=4$, since only the ratio of the coupling parameters appear in the subsequent calcuations. The 
most exciting property of this model is that even in two dimensions  it can be exactly solved using Jordan-Wigner (JW)
 transformation\ct{lieb61,barouch70,kogut79,bunder99,kitaev06,chen,feng}
in terms of Majorana fermions  given by
\ba
&& a_{j,l}=\left(\prod_{i=-\infty}^{j-1}\sigma_{i,l}^z\right)\sigma_{j,l}^y \hspace{2mm}\text{for even $j+l$,}\nonumber\\
&& a'_{j,l}=\left(\prod_{i=-\infty}^{j-1}\sigma_{i,l}^z\right)\sigma_{j,l}^x\hspace{2mm}\text{for even $j+l$,}\nonumber\\
&& b_{j,l}=\left(\prod_{i=-\infty}^{j-1}\sigma_{i,l}^z\right)\sigma_{j,l}^x\hspace{2mm}\text{for odd $j+l$,}\nonumber\\
&& b'_{j,l}=\left(\prod_{i=-\infty}^{j-1}\sigma_{i,l}^z\right)\sigma_{j,l}^y\hspace{2mm}\text{for odd $j+l$.} 
\label{majoranafermions}
\ea
Here,  $a_{j,l}$, $a'_{j,l}$, $b_{j,l}$ and $b'_{j,l}$ are all Majorana fermion operators, they obey the 
relations $a^{\dagger}_{j,l}=a_{j,l}, b^{\dagger}_{j,l}=b_{j,l}, 
\{a_{j,l},a_{m,n}\}= \{b_{j,l},b_{m,n}\}= 2\delta_{j,m}\delta_{l,n}$ and $\{a_{j,l},b_{m,l}\}=0$. One can now change the lattice site indices $\left(j,l\right)$ 
 of honeycomb lattice to a 2-dimensional vector $\vec{n}$ , where $\vec{n}=\sqrt{3}\hat{i}n_1+
\left(\frac{\sqrt{3}}{2}\hat{i}+\frac{3}{2}\hat{j}\right)n_2$ which labels the midpoints of the vertical bonds of the honeycomb lattice. Here $n_1$ and $n_2$
take all integer values 
 so that the vectors $\vec{n}$ form a triangular lattice. The Majorana fermions $a_{\vec{n}}$ and $b_{\vec{n}}$ are placed at the top and bottom 
sites respectively 
of the bond labeled by $\vec{n}$. The whole lattice is spanned by the vectors $\vec{M_1}= \frac{\sqrt{3}}{2}\hat{i}-\frac{3}{2}\hat{j}$
 and $\vec{M_2}= \frac{\sqrt{3}}{2}\hat{i}+\frac{3}{2}\hat{j}$, see Fig.~(\ref{Fig1}). 

Under the  transformation to Majorana fermions as defined in Eqs.~(\ref{majoranafermions}), Hamiltonian (\ref{ham1}) takes the form
\be
H=i\sum_{\vec{n}}\left(J_1b_{\vec{n}}a_{\vec{n}-\vec{M_1}}+J_2b_{\vec{n}}a_{\vec{n}+\vec{M_2}}+J_3D_{\vec{n}}b_{\vec{n}}a_{\vec{n}}\right),
\label{ham2} 
\ee
where $D_{\vec{n}}$=$ib'_{\vec{n}}a'_{\vec{n}}$. These $D_{\vec{n}}$ operators have eigenvalues $\pm1$ independently for each $\vec{n}$ and commute with each other 
and also with $H$ which makes the Kitaev model exactly solvable. Since $D_{\vec{n}}$ is a constant of motion one can use one of the eigenvalues $\pm1$ for each 
$\vec{n}$ in the Hamiltonian. The ground state of the model corresponds to $D_{\vec{n}}=1~ \forall~ \vec{n}$~ \ct{kitaev06}.
With  $D_{\vec{n}}=1$,  we can easily diagonalize the Hamiltonian (\ref{ham2}) quadratic in Majorana fermions.

\begin{figure}[ht]
\begin{center}
\includegraphics[width=7.7cm]{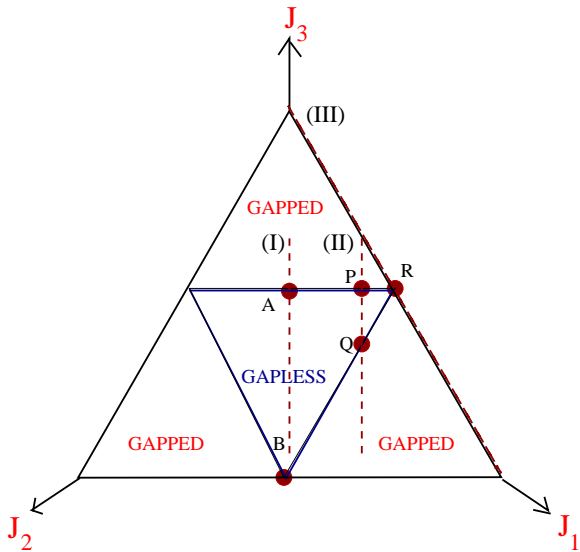}
\end{center}
\caption{(Color online) Figure shows phase diagram of the Kitaev model, satisfying  $J_1+J_2+J_3=4$. The inner equilateral triangle corresponds to the 
gapless phase in which the coupling parameters satisfies the relations $J_1\leq J_2+J_3$, $J_2\leq J_3+J_1$ 
and $J_3\leq J_1+J_2$. Along the three paths I, II and III $J_3$ is varied, so as to
 study the LE. The path I, II and III are defined by the equations $J_1=J_2$, $J_1=J2+1$ and $J_1+J_3=4$ respectively.}
\label{Fig2}
\end{figure}

The Fourier transform of the Majorana fermions can be defined as
\be
a_{\vec{n}}= \sqrt{\frac{4}{N}}\sum_{\vec{k}}\left(a_{\vec{k}}e^{i\vec{k}.\vec{n}}+a_{\vec{k}}^{\dagger} e^{-i\vec{k}.\vec{n}}\right),
\label{ftransform}
\ee
similarly $b_{\vec{n}}$ also has same Fourier transform relation. The $a_{\vec{k}}$'s and $b_{\vec{k}}$'s are Dirac fermions which follow the fermionic 
anti-commutation 
relations. Here, $N$ is the total number of sites and $N/2$ is the number of unit cells. In the above sum given in Eq.~(\ref{ftransform}), $\vec{k}$ is
 extended over half of the 
Brillouin zone of the hexagonal lattice due to Majorana nature of the fermions \ct{sengupta08}. We recall that the full Brillouin 
zone on the reciprocal lattice represents a rhombus with 
vertices $\left(k_x,k_y\right)$ = $\left(\pm2\pi\sqrt{3},0\right)$ and $\left(0,\pm2\pi/3\right)$. In the momentum space the Hamiltonian (\ref{ham2}) takes the form
$H=\sum_{\vec{k}}\psi^{\dagger}_{\vec{k}}H_{\vec{k}}\psi_{\vec{k}}$ where $\psi^{\dagger}_{\vec{k}}=\left(a^{\dagger}_{\vec{k}},b^{\dagger}_{\vec{k}}\right)$ and 
the reduced $2 \times 2$ Hamiltonian
 $H_{\vec{k}}$, can be expressed  in terms of Pauli matrices as

\ba
&& H_{\vec{k}}= \alpha_{\vec{k}}\sigma^1 + \beta_{\vec{k}}\sigma^2,\nonumber\\ 
\text{where} \hspace{2mm} && \alpha_{\vec{k}} = 2[J_1\sin(\vec{k}.\vec{M_1})-J_2\sin(\vec{k}.\vec{M_2})],\nonumber\\
\text{and}\hspace{1mm} && \beta_{\vec{k}} = 2[J_3+J_1\cos(\vec{k}.\vec{M_1})+J_2\cos(\vec{k}.\vec{M_2})].
\label{ham3}
\ea
The eigenenergies of the $H_k$ are given by
\be
 E^{\pm}_{\vec{k}}= \pm \sqrt{\alpha^2_{\vec{k}}+\beta^2_{\vec{k}}}.
\label{energy1}
\ee
This energy spectrum corresponds to two energy bands; it is noteworthy that for $|J_1-J_2|\leq J_3\leq(J_1+J_2)$, the band
 gap $\Delta_{\vec{k}}= E^+_{\vec{k}} - E^-_{\vec{k}}$ vanishes 
for some particular $\vec{k}$ modes leading to the gapless phase of the Kitaev model. 
The phase diagram 
of the model is shown in  an equilateral triangle  satisfying the relation $J_1+J_2+J_3=4$ and 
$J_1,J_2,J_3>0$ (see Fig.~(\ref{Fig2})); one can easily show that the whole phase is divided into
three gapped phases, separated by a gapless phase (inner equilateral triangle) which is   
bounded by gapless critical lines 
$J_1=J_2+J_3$, $J_2=J_3+J_1$ and $J_3=J_1+J_2$. 

 On the critical line 
$J_3=J_1+J_2$ energy gap goes to zero for the four $\vec{k}$ modes given by $(k_x,k_y)$= $(\pm2\pi/\sqrt{3},0)$ and $(\pm2\pi/3,0)$ which are the 
four corner points of Brillouin zone. One can now expand $\alpha_{\vec{k}}$ and $\beta_{\vec{k}}$ around one of the critical modes for $J_3=J_1+J_2$, in the form
\ba
&& \alpha_{\vec{k}}=  \sqrt{3}(J_2-J_1)k_x + 3(J_1+J_2)k_y,\nonumber\\
&& \beta_{\vec{k}}= \frac{3}{4}(J_1+J_2)k^2_x + \frac{9}{4}(J_1+J_2)k^2_y + \frac{3\sqrt{3}}{2}(J_2-J_1)k_xk_y,\nonumber\\
\label{alphabeta1} 
\ea
where $k_x$ and $k_y$ are the deviations from the above mentioned critical modes.  We note that  $\alpha_{\vec{k}}$ varies linearly
and  $\beta_{\vec{k}}$ varies  quadratically  
 in $k_x$ and $k_y$.  The point $J_{3c}=2J_1$ (where $J_1=J_2$) denoted by A in (Fig.~(\ref{Fig2})) needs to be checked
 carefully. This is an AQCP \ct{hikichi} with energy dispersion
 $E_{\vec{k}}\sim k^2_x$ along $k_x$  
 ($k_y=0$) and $E_{\vec{k}}\sim k_y$ along $k_y$ ($k_x=0$).
The corresponding dynamical exponents are given by $z_{\perp}=1$ and $z_{\parallel}=2$, respectively.

For $J_1\neq J_2$, $J_{3c}=J_1+J_2$ is  also an AQCP that can be shown using a rotation to
a new coordinate system \ba
&& k_1= \sqrt{3}(J_2-J_1)k_x + 3(J_1+J_2)k_y,\nonumber\\
&& k_2= 3(J_1+J_2)k_x - \sqrt{3}(J_2-J_1)k_y, \label{k1k2}
\ea
in which $\al_{\vec k}$ and $\beta_{\vec k}$ take the form
\ba
&& \alpha_{\vec{k}}= k_1,\nonumber\\ 
&& \beta_{\vec{k}}= c_1k^2_1 + c_2k^2_2 + c_3k_1k_2,
\ea
where  $c_1 = 9(J_1+J_2)(4J^2_1+4J^2_2+J_1J_2)$,
$c_2 = 27J_1J_2(J_1+J_2)$, and 
$ c_3 = 18\sqrt{3}J_1J_2(J_2-J_1)$ .
Therefore, for a general AQCP $(J_1\neq J_2)$ the dispersion will vary linearly and quadratically along $\hat{k_1}$ and $\hat{k_2}$ directions, respectively, with two 
dynamical exponents $z_1=1$ and $z_2=2$.  

\section{Qubit Coupled To $J_{3}$ Term Of the Kitaev Hamiltonian}
\label{III}
In this section we will provide a general calculation of the LE considering Kitaev model on a honeycomb lattice as an environment ($E$) that is coupled to a 
central spin-$\frac{1}{2}$ ($S$).
 We shall denote the ground state and excited state of the central spin $S$
by $\ket{g}$ and $\ket{e}$ respectively. $S$ is coupled to $J_3$ term of $E$ Hamiltonian only when the central spin is in the excited state 
$\ket{e}$. Therefore the composite Hamiltonian takes the form
\ba
H_T\left(J_3,\delta\right) &=& \sum_{j+l=even}(J_{1}\sigma^{x}_{j,l}\sigma^{x}_{j+1,l}+ J_{2}\sigma^{y}_{j-1,l}\sigma^{y}_{j,l}\non\\
&+& J_{3}\sigma^{z}_{j,l}\sigma^{z}_{j,l+1}+\delta\ket{e}\bra{e}\sigma^{z}_{j,l}\sigma^{z}_{j,l+1}),
\label{ham4}
\ea
where $\delta$ is the coupling strength of $S$ to $E$. We shall work in the limit of $\de\rightarrow0$. 

We consider  that the $S$ is initially in a generalized state $\ket{\phi(0)}_S=c_g\ket{g}+c_e\ket{e}$ (with the coefficients satisfying the condition 
$|c_g|^2+|c_e|^2=1$), and the $E$ is  initially  in the ground state $\ket{\varphi(J_3,0)}$. The evolution of the environmental spin model 
splits into two branches, given by $\ket{\varphi(J_3,t)}= \exp(-iH(J_3)t)\ket{\varphi(J_3,0)}$ and 
$\ket{\varphi(J_3+\delta,t)}= \exp(-iH(J_3+\delta)t)\ket{\varphi(J_3,0)}$;  the evolution of $\ket{\varphi(J_3,t)}$ is driven by the 
Hamiltonian $H(J_3)=H_T(J_3,0)$ (when the $S$ is in the ground state and hence there is no $\delta$ term present in the Hamiltonian),
whereas $\ket{\varphi(J_3+\delta,t)}$ evolves with $H(J_3+\delta)=H_T(J_3,0)+V_e$, where $V_e= \delta\sum_{j+l=\text{even}}\sigma^z_{j,l}\sigma^z_{j,l+1}$,
 is the effective potential arising due to the coupling between $S$ and $E$.  The wave function of the composite system 
at a time $t$ is given by
\be
\ket{\psi(t)}= c_g\ket{g}\otimes\ket{\varphi(J_3,t)} + c_e\ket{e}\otimes\ket{\varphi(J_3+\delta,t)}.
\label{twf} 
\ee
As a result the LE is given by
\ba
 L\left(J_3,t\right) &=& |{\langle\varphi(J_3,t)|\varphi(J_3+\delta,t)\rangle}|^2,\nonumber\\ 
 &=& |{\langle \varphi(J_3,0)|\exp(-iH(J_3+\delta)t)|\varphi(J_3,0)\rangle}|^2.\label{le}
\ea
Here, we have exploited the fact that the $\ket{\varphi(J_3,0)}$ is an eigenstate of the Hamiltonian $H(J_3)$.

Following Fourier transformation and Bogoliubov transformation the diagonalized form of the Hamiltonian (\ref{ham1}) is given by
\be
H(J_3) = \sum_{\vec{k}}[-\varepsilon_{\vec{k}}(J_3)A^{\dagger}_{\vec{k}}A_{\vec{k}} + \varepsilon_{\vec{k}}(J_3)B^{\dagger}_{\vec{k}}B_{\vec{k}}],
\label{ham6} 
\ee
where the $A_{\vec{k}}$'s and $B_{\vec{k}}$'s are Bogoliubov fermionic operators defined as
\ba
&& A_{\vec{k}}=\frac{1}{\sqrt{2}}[a_{\vec{k}} - e^{-i\theta_{\vec{k}}}b_{\vec{k}}],
\hspace{1mm} B_{\vec{k}}=\frac{1}{\sqrt{2}}[a_{\vec{k}} + e^{-i\theta_{\vec{k}}}b_{\vec{k}}],\nonumber\\
\text{with}\hspace{1mm}&& e^{i\theta_{\vec{k}}}= \frac{\alpha_{\vec{k}}+i\beta_{\vec{k}}}{\sqrt{\alpha^2_{\vec{k}}+\beta^2_{\vec{k}}}},\label{boperators} 
\ea
 and the
energy spectrum is given by (see Eqs.(\ref{ham3}) and (\ref{energy1}))
\be
\varepsilon_{\vec{k}}(J_3)= \sqrt{\alpha^2_{\vec{k}} + \beta^2_{\vec{k}}}
\hspace{2mm}\text{and}\hspace{2mm} \varepsilon_{\vec{k}}(J_3+\delta)= \sqrt{\alpha^2_{\vec{k}} + \beta'^2_{\vec{k}}},
\label{energy2}
\ee
where $\alpha_{\vec{k}}$ and $\beta_{\vec{k}}$ are defined in Eq.(\ref{ham3}), and $\beta'_{\vec{k}}$ corresponds to
 the value with $J_3+\delta$ instead of $J_3$. 
The complete ground state of $H(J_3)$ can be written in the form (see Ref. [\onlinecite{victor12}] for details),
\be
\ket{\varphi(J_3,0)} = \prod_{\vec{k}}[\frac{1}{2}(a^{\dagger}_{\vec{k}}-e^{i\theta_{\vec{k}}}b^{\dagger}_{\vec{k}})\hspace{2mm}(a'^{\dagger}_{\vec{k}}+ib'^{\dagger}_{\vec{k}})]\ket{\Phi} 
\label{gstate}
\ee
where $\vec{k}$ runs over half of the Brillouin zone of the hexagonal lattice. Following mathematical
steps identical to those in [\onlinecite{zanardi,sharma12}], it can be shown that Eq.(\ref{gstate}) leads to the expression for the LE  given by
\be
L(J_{3},t)=\prod_{\vec{k}}L_{\vec{k}} = \prod_{\vec{k}}[1-\sin^2(2\phi_{\vec{k}})\sin^2(\varepsilon_{\vec{k}}(J_3+\delta) t)], \label{le1}
\ee
where, $\tan\theta_{\vec{k}}(J_3+\delta)= \alpha_{\vec{k}}/\beta'_{\vec{k}}$ and $\phi_{\vec{k}}= [\theta_{\vec{k}}(J_3) - \theta_{\vec{k}}(J_3+\delta)]/2$.
The expression for  LE closely resembles that of  the case when the transverse Ising chain is chosen to be the environment \ct{zanardi}. For 
numerical analysis of Eq.(\ref{le1}), we shall use $k_x$ and $k_y$ in terms of two 
independent variables $v_1$ and 
$v_2$, with $0\leq v_1,v_2\leq 1 $. The $k_x$ and $k_y$ are given by \ct{sengupta08}
\be
k_x=\frac{2\pi}{\sqrt{3}}(v_1+v_2-1), \hspace{3mm} k_y=\frac{2\pi}{3}(v_2-v_1),
\label{kxky} 
\ee
which span the rhombus uniformly. Avoiding the corner points of the Brillouin zone ( where the LE results in a  zero value), we 
vary $v_1$ and $v_2$ from $1/(2N)$ to $1-1/(2N)$ in steps of $1/N$, where $N$ is the system size \ct{victor12} and consider only the half of the Brillouin
 zone using the condition $(v_1, v_2)\geq 1$.
  
\begin{figure}[ht]
\begin{center}
\includegraphics[width=7.7cm]{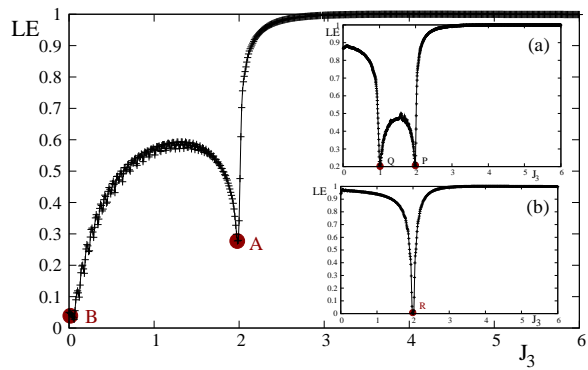}
\end{center}
\caption{ LE as a function of parameter $J_3$ ($J_3$ is varied along path 'I') shows a sharp dip at point A ($J_3=2-\delta$) and after a small revival in the gapless phase it again decays at ponit B,
 $J_3=0$ (see Fig.~(\ref{Fig2})) with $N_x=N_y=200$, $\de=0.01$ and 
$t=10$. Inset (a) shows the variation in LE when the parameter $J_3$ is varied along the path II ($J_1=J_2+1$) in the phase diagram for 
$N_x=N_y=200$, $\de=0.01$ and 
$t=10$ clearly showing a sharp dip at point P ($J_3=2-\de$) and again rise at point Q ($J_3=1-\de$).  
 Inset (b) marks the dip in LE when $J_3$ is varied along the path III ($J_1+J_3=4$), for this case $N=400$, $\de=0.01$ and 
$t=10$ so that $E$ realise the change in the behavior at $J_3=2-\de$. Details of these three cases  is provided in the subsection (\ref{a}), (\ref{b})
 and (\ref{c}) respectively. }
\label{Fig3}
\end{figure}

The LE is calculated numerically as a function of $J_3$ using Eq.(\ref{le1}) and it shows dip
at all critical points. To illustrate this, we choose three paths along which the interaction $J_3$ is varied.  In the first case  $J_3$ is varied along the 
path $J_1=J_2$ (path `I' in Fig.~(\ref{Fig2})) so that the model  enters from the gapped phase to  the gapless phase (extending in the region  $J_3\in[0,2]$) 
crossing the AQCP (point `A' in Fig.~(\ref{Fig2})) at 
 $J_3=2-\delta$.  The LE shows a sharp dip at point A and there is a revival with a small magnitude which again decays at the end
 point B, $J_3=0$ (see Fig.~(\ref{Fig3})). Now surprising result shows up when the path is so chosen (path `II' in Fig.~(\ref{Fig2}), given
 by the equation $J_1=J_2+1$)  
 that the system enters the gapless phase through an AQCP with $J_1 \neq J_2$, (denoted by `P' in Fig.~(\ref{Fig2})). The LE 
 shows a sharp dip at the point P and stays close to its minimum value (with a small revival as 
 observed in path I) throughout the gapless phase and 
 again shows a rise when the system exits the gapless phase through the point Q.
 In contrary, for the case when $J_3$  is changed along the line $J_1+J_3=4$, $J_2 =0$ (path `III' in Fig.~(\ref{Fig2})), one
 observes only a single drop in the LE near $J_3=2-\delta$ (see Fig.~(\ref{Fig3}), inset (b)); this is associated with the critical point of the
 one-dimensional Kitaev model. 
 In the next section 
we will study the scaling of the short time behaviour of LE close to these critical points and the collapse and reviaval 
of LE with time when the $E$ is right at the critical point.

\subsection{Path I: Anisotropic Quantum Critical Point ($J_1=J_2$)}
\label{a}
As discussed in Sec.\ref{II}, $J_{3c}=2J_1$ is an AQCP with critical exponents $\nu_{\perp}=z_{\perp}=1$ along $\hat j$ direction and 
$\nu_{\parallel}=1/2,z_{\parallel}=2$ along 
$\hat i$ direction. At this point energy gap vanishes for the three critical modes  given by $(2\pi/\sqrt{3},0)$ and $(0,\pm 2\pi/3)$ in half 
of the Brillouin zone. Now 
we will study the short time behavior of the LE (in Eq.~(\ref{le1})) close to the AQCP. 
We define a cutoff frequency $K_c=\left(K_{x,c},K_{y,c}\right)$ such that modes up to this cut-off only are
considered to calculate the decay of LE at short time close to the AQCP. Then the LE is given by
\be
L_c\left(J_3,t\right)=\prod_{k_x,k_y>0}^{K_c}L_{\vec{k}}.
\label{lc1} 
\ee
We define the quantity $\mathcal{S}(J_3,t)$, such that $\mathcal{S}\left(J_3,t\right)=\ln L_c\equiv-\sum_{k_x,k_y>0}^{K_c}|\ln L_{\vec{k}}|$. Expanding 
around one of the critical mode upto the cut-off, we get 
 $\sin^2\varepsilon_k(J_3+\delta)t \approx 4\left(J_3+\delta-2J_1\right)^2t^2$ and $\sin^2\left(2\phi_k\right)\approx9J_1^2k_y^2\delta^2/\left(J_3-2J_1\right)^2
\left(J_3+\delta-2J_1\right)^2$ therefore we obtain,
\be
\mathcal{S}\left(J_3,t\right) \approx -\frac{36\mathcal{E}\left(K_c\right)J_1^2\delta^2t^2}{\left(J_3-2J_1\right)^2},
\label{sj3} 
\ee
where $\mathcal{E}\left(K_c\right)$=$4\pi^2N_c\left(N_c+1\right)\left(2N_c+1\right)/54N_y^2$ and $N_c$ is a integer nearest to $3N_yK_c/2\pi$. 
We therefore find an exponential decay of the LE in the early time limit given by
\be
L_c\left(J_3,t\right)\approx \exp\left(-\gamma t^2\right)
\label{lc2} 
\ee
where $\gamma=36\mathcal{E}\left(K_c\right)J_1^2\delta^2/\left(J_3-2J_1\right)^2$. 
The anisotropic nature of the quantum critical point is reflected in the fact that $\gamma$ scales as $1/N_y^2$ and is independent of $N_x$. 
Further using the expression of $L_c\left(J_3,t\right)$, one can easily observe that it is invariant under the transformation $N_y\rightarrow N_y \alpha, \delta\rightarrow \delta/\alpha$ and $t\rightarrow t\alpha$, where $\alpha$ is some integer. 

\begin{figure}[ht]
\begin{center}
\includegraphics[width=8.7cm, height=6.5cm]{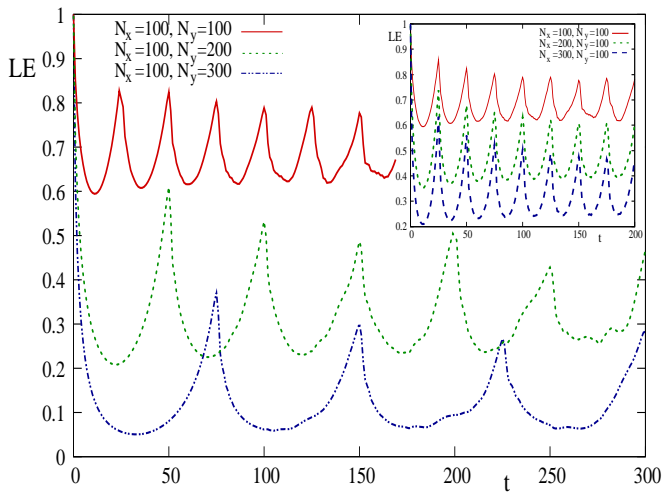}
\end{center}
\caption{ (Color online) The collapse and revival of LE with t at the AQCP (point `A' in Fig.~(\ref{Fig2})) for $J_1=J_2=1$, $\de=0.01$ and 
$J_3=2-\delta$, keeping $N_x (=100)$, fixed and varying $N_y$ verifies
 the scaling relations satisfied by $N_y$, $\de$ and $t$ as discussed in the text. The inset shows that the quasiperiod of the collapse and revival is independent of $N_x$. }
\label{Fig4}
\end{figure}

Now we fix $J_1=J_2=1$ and $J_3=2-\delta$ (point `A' in Fig.~(\ref{Fig2})) and observe collapse and revival of LE with time (presented in the Fig.~(\ref{Fig4})). The time period of collapse and revival is proportional to $N_y$, 
and is unaffected by the changes in $N_x$; this confirms the scaling result of the decay rate $\gamma$ for the short time limit near the AQCP discussed above.

\subsection{Path II: Anisotropic Quantum Critical Point $\left(J_1 \neq J_2\right)$}
\label{b}
It has been shown that the point P in Fig.~(\ref{Fig2}) ($J_1\neq J_2$, $J_{3,c}=J_1+J_2 $)
is an AQCP which can be seen by choosing directions
$\hat{k_1}= \sqrt{3}\left(J_2-J_1\right)\hat{i} + 3\left(J_1+J_2\right)\hat{j} $ (see Sec \ref{II}) and $\hat{k_2}$, perpendicular
 to $\hat{k_1}$ \ct{victor12}. 
The critical exponents associated with this critical point are given by $\nu_1=z_1=1$, and $\nu_2=1/2,z_2=2$ along $\hat k_1$ and $\hat k_2$ directions, respectively. 
To calculate the early time scaling in a similar spirit as in the previous section, 
we expand Eq.~(\ref{le1}) near one of the critical modes up to the cut-off $K_c$ to  obtain 
$\sin^2\left(\varepsilon_k(J_3+\delta)\right)t \approx 4\left(J_3+\delta-J_1-J_2\right)^2t^2$ and $\sin^2\left(2\phi_k\right)\approx k_1^2\delta^2/
4\left(J_3-J_1-J_2\right)^2\left(J_3+\delta-J_1-J_2\right)^2$. In the short time limit, the LE becomes
\be
L_c\left(J_3,t\right) \approx \exp\left(-\gamma t^2\right)
\label{lc3} 
\ee
where, $\gamma=\delta^2\mathcal{E}\left(K_c\right)/\left(J_3-J_1-J_2\right)^2$, $\mathcal{E}\left(K_c\right)= 8\pi^2J_2^2N_c\left(N_c+1\right)\left(2N_c+1\right)/3N^2$ and
$N_c$ is an integer nearest to $NK_c/4\pi J_2$.

\begin{figure}[ht]
\begin{center}
\includegraphics[width=7.7cm]{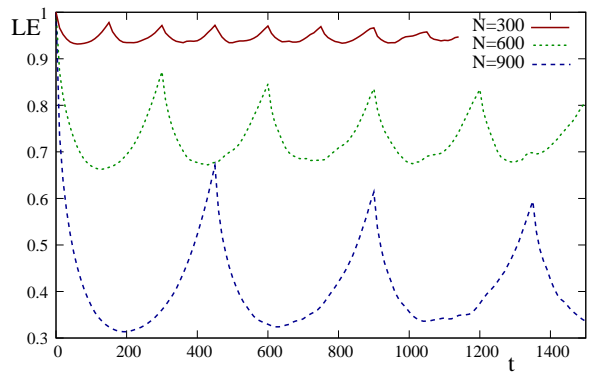}
\end{center}
\caption{ (Color online) Variation of LE with $t$ at the AQCP P ($J_1=3/2,J_2=1/2$ and $J_3=2-\de$) shows collapse 
and revival with diffrent $N_x=N_y=N$ and $\de=0.001$. }
\label{Fig5}
\end{figure}

In fact comparing with the previous section \ref{a}, one can see that in this case $k_1$ (instead of $k_y$) appears in
the expression  of the LE in the short-time limit. 
 Further, from
 Eq. (\ref{lc3}) and the expression of $\gamma$ one observes that $L_c\left(J_3,t\right)$ is invariant under the 
transformation $N_x=N_y=N\rightarrow N\alpha,\delta\rightarrow\delta/\alpha$ and 
$t\rightarrow t\alpha$, with $\alpha$ being some integer 
which is also observed in the collapse and revival behavior (see Fig.~(\ref{Fig5})).

\subsection{Path III: One-dimensional Quantum Critical Point $\left(J_2=0\right)$}
\label{c}
As mentioned already, along the line $J_1+J_3=4$,  ($J_2=0$), the two dimensional spin model reduces to an equivalent one dimensional  spin chain with energy gap vanishing at $J_1=J_3$ for  
$k_c=\pi$ and the corresponding dynamical exponent being $z=1$. We shall now expand $\sin\left(\varepsilon_k(J_3,\delta)t\right)$ and $\sin\left(2\phi_k\right)$ around the critical mode $k_c$ to analyze the short time 
decay of LE, resulting into $\sin^2\left(\varepsilon_k(J_3+\delta)t\right)\approx 4\left(J_3+\delta-J_1\right)^2t^2$ and 
$\sin^2\left(2\phi_k\right)\approx J_1^2k^2\delta^2/\left(J_3-J_1\right)^2\left(J_3+\delta-J_1\right)^2$. The LE  hence takes the from
\be
L_c\left(J_3,t\right)\approx \exp\left(-\gamma t^2\right)
\label{lc4} 
\ee
\begin{figure}[ht]
\begin{center}
\includegraphics[width=7.7cm]{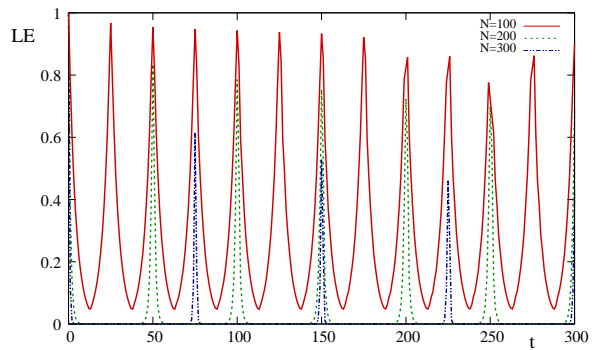}
\end{center}
\caption{ (Color online) The LE as function of time at the QCP (point `R' in Fig.~(\ref{Fig2})) $J_1=2$ and $J_3=2-\de$ with different N and $\de=0.01$, verifying the analytical 
scaling with $N$, $\de$ and $t$. }
\label{Fig6}
\end{figure}
where, $\gamma=4J_1^2\delta^2\mathcal{E}\left(K_c\right)/\left(J_3-J_1\right)^2$ and $\mathcal{E}\left(K_c\right)= 4\pi^2N_c\left(N_c+1\right)\left(2N_c+1\right)/6N^2$ , $N_c$ is an integer nearest to 
$NK_c/2\pi$. From Eq.~(\ref{lc4}), we find  that the LE shows a similar scaling relation as is
expected for a one-dimensional chain with $z=1$ \ct{zanardi};  this is also confirmed by studying the 
collapse and revival of LE (see Fig.~(\ref{Fig6})).

\section{Conclusion}
In this paper we study a variant of the central spin model in which a central spin (qubit) is globally coupled to an environment which is chosen to be a 
two-dimensional  Kitaev model on a honeycomb lattice through the interaction term $J_3$. Using the exact solvability of the Kitaev model,
we have derived  an exact expression of the LE when the interaction $J_3$ is varied in a way such that
the system enters the gapless phase crossing the AQCP of the phase diagram. However, the 
behavior of the LE as a function of $J_3$ depends upon the path along which $J_3$ is varied. In the
case when the AQCP, Q (with $J_1 \neq J_2$ see Fig.~(\ref{Fig2})) is crossed, one observes a complete revival of the echo when 
the system exits the gapless phase to re-enter the gapped phase; this is in contrast
to the case $J_1 =J_2$. For the case of $J_2=0$ there is only one sharp dip at the critical point $J_3=2-\delta$ which is associated 
with the QCP of the one-dimensional Kitaev model. 

The early time scaling behavior for both the paths I and II close to the AQCP bear the signature
of the fact that the gapless phase is entered crossing an AQCP with different exponents along
different spatial directions. This is also confirmed by studying the collapse and the revival of
the LE as a function of time. However, one does not observe a perfect collapse and revival (except
for the equivalent one dimensional case); this may be because of the proximity to a gapless phase. The quasi-period of collapse and revival
 in all cases scale with the system size as $N^z$.  The case with $J_2=0$ reflects the fact that the system is essentially
one-dimensional in this limit.  It is straightforward to relate these results to the decoherence
of the central spin close to a critical point.

{This study of LE can  be verified experimentally as presented by Zhang $et~al$ \ct{zhang09}; they measure the LE as an indicator
 of quantum criticality for a
one-dimensional quantum Ising model with an antiferromagnetic interaction using NMR quantum
simulators. In this experiment, they prepare the ground state of the Hamiltonian (using the gate sequences) which need not 
be the true ground state but could be a state that
approximates the ground state of the system well, and then measure the LE for finite number of spins. Similar experiments can be realized with 
the approximate ground state of the Kitaev model. Also, since the Kitaev model can also be realized using an optical lattice\ct{duan03, sen08} 
(where the couplings can be separately tuned with
the help of different microwave radiations), there exists a possibility of verifying these results in an optical lattice also.}

{It should be noted that  in a recent work, Pollmann $et~al$ \ct{pollmann10}, have studied the problem of the LE in a transverse 
Ising spin chain in the presence 
of a longitudinal field; more precisely they calculated
the magnitude of the overlap between the final state reached following a slow quench across the 
QCP and its time evolved counterpart at time $t$ (generated following the time evolution with the final Hamiltonian). They 
observe a cusp-like minimum in the echo as a function of time in the
limit when the spin chain is integrable. However, this behavior is smeared in the non-integrable
case (with non-zero longitudinal field) thus providing a probe for integrable versus non-integrable
behavior. In the present paper, we  however deal with an equilibrium situation in which the spin
chain is not quenched across the QCP, and observe the collapse and revival only at the QCP.}

\begin{center}
\bf{Acknowledgements}
\end{center}
We acknowledge  Amit Dutta, Victor Mukherjee and Aavishkar Patel for helpful discussions and comments. AR acknowledges B. K. 
Chakrabarti for valuable discussions and thanks IIT Kanpur for financial support during this work. SS thanks CSIR, New Delhi for Junior Research Fellowship.




\end{document}